\documentclass{jltp}

\usepackage{latexsym}
\usepackage{amssymb}
\usepackage{amsmath}
\usepackage{epsfig}
\usepackage{amsthm}

\newcommand{\vac}{|\text{vac}\rangle}

\title{Exact stationary state of a staggered stochastic hopping 
model{\footnote{Dedicated to Prof. Peter W\"olfle on the
occasion of his sixtieth birthday.}}}

\author{Kai Klauck, Andreas Schadschneider and Johannes Zittartz}

\address{Institut f\"ur Theoretische Physik, Universit\"at zu K\"oln,
Z\"ulpicher Str.~77,\\
D-50937 K\"oln, Germany}

\runninghead{K.~Klauck, A.~Schadschneider and J.~Zittartz}{Exact stationary 
state of a staggered hopping model}

\begin{document}

\maketitle

\begin{abstract}
We determine the $N$-particle stationary states of a staggered stochastic
hopping model with reflective boundaries. It is shown that the 
stationary states are in fact so-called optimum ground states. 
Recursion relations in the particle number for any
$l$-point density correlation function will be derived. Furthermore, 
the connection between reflective boundaries and the
occurrence of optimum ground states is examined. An explicit
counterexample shows that reflective boundaries do not
enforce the stationary state to be an optimum ground state.

\end{abstract}


\section{INTRODUCTION}

Interacting many-particle systems encountered in nature are sometimes
difficult to describe in terms of classical or quantum mechanics.  
A stochastic description is then more appropriate for systems which
behave essentially randomly on a phenomenological level.  Often this
leads to problems far away from thermal equilibrium 
for which -- in contrast to equilibrium systems -- a general 
concept or framework is still missing. 
Therefore one has to rely on the investigation of special examples. 
Models which are exactly solvable\cite{schuetzdomb} become then
even more important than in equilibrium physics\cite{baxter}.
They allow to probe the nature of non-equilibrium physics with the
problems related to the reliability of approximative methods.
Also stochastic systems are often used for the modelling of interdisciplinary
problems\cite{stauffer,stanley,bouchaud,review} where the interactions 
between the entities are not known exactly. 

The dynamics of stochastic systems is governed by the master equation
which can be interpreted as a Schr\"odinger equation in imaginary
time\cite{alcaraz} (see Sec.~\ref{sec_def}). In the analytical
treatment of non-equilibrium systems it is therefore possible to
exploit this similarity by using successful techniques from 
equilibrium physics. Especially the so-called matrix product 
Ansatz\cite{evans} (MPA), first developed for spin chains\cite{mpa}, 
is important. It has been shown\cite{ks,us} that the stationary state 
of one-dimensional stochastic processes is generically of MPA-type.
However, in most cases it is still very difficult to calculate expectation
values explicitly.

In this paper the exact stationary state of a one-dimensional 
stochastic process with staggered interaction and reflective 
boundaries is determined. This model can be regarded as the merger 
of the $U_q[SU(2)]$-symmetric hopping model of Sandow and 
Sch\"utz\cite{sandowschuetz} and a ratchet model proposed by Kolomeisky and 
Widom\cite{kolo} (for details, see \cite{klauck}). A similar stochastic 
hopping model with periodic boundary conditions, where also coagulation 
and decoagulation processes may occur, has been studied by Fujii
\cite{fuji} using free fermion techniques. Our model can not be solved
using free fermion techniques but requires a different approach. This
approach relies heavily on the observation that the stationary state
of our model is an {\it optimum ground state} (OGS) \cite{deboer,zitt2,nigge}.
The occurrence of OGS in the context of matrix product states will be
elucidated. It will be shown that OGS are not ubiquitous
for systems with reflective boundaries.


\section{QUANTUM FORMALISM AND DEFINITION OF THE MODEL}
\label{sec_def}

We consider stochastic systems on a one-dimensional lattice with $L$ sites.
Each cell $j$ can either be empty or occupied by a particle. Its state
is characterized by the variable $\alpha_j$ which could e.g.\ be the
occupation number. The model is then defined by its dynamics, namely
by the transition rates $w(\alpha\to\beta)$ for a transition from
a lattice configuration $\alpha=(\alpha_1,\ldots,\alpha_L)$ to 
the configuration $\beta=(\beta_1,\ldots,\beta_L)$.
The probability $P(\alpha;t)$ to find the system at time $t$ in the
configuration $\alpha=(\alpha_1,\ldots,\alpha_L)$ is then given
by the master equation
\begin{equation}
\frac{\partial P(\alpha;t)}{\partial t} = \sum_{\{\beta\}} 
w(\beta\to\alpha) P(\beta;t) - \sum_{\{\beta\}} 
w(\alpha\to\beta)P(\alpha;t).
\label{master}
\end{equation}

Introducing the state vector $|P(t)\rangle=\sum_{\alpha} P(\alpha_1,\ldots,
\alpha_L;t)|\alpha_1,\ldots,\alpha_L\rangle$ the master equation can
be written in more compact form
\begin{equation}
\label{schroed}
\frac{\partial}{\partial t} |P(t)\rangle =-{H} \, |P(t)\rangle ,
\end{equation}
where the matrix elements of the stochastic Hamiltonian $H$ can
be expressed through the transition rates $w(\beta\to\alpha)$.
For processes with nearest-neighbour interaction and periodic
boundary conditions, the stochastic Hamiltonian $H$ can be written
as a sum $H=\sum_{j=1}^L h_{j,j+1}$  of local Hamiltonians $h_{j,j+1}$ 
acting only on the sites $j$ and $j+1$.
For other boundary conditions the interactions at the boundaries
$j=1$ and $j=L$ are described by additional boundary operators
$h_1$ and $h_L$.

The so-called quantum formalism described above emphasizes the 
similarity with a Schr\"odinger equation in imaginary time.
The stochastic Hamiltonian $H$, which in general is not hermitean,
is often related to spin chains\cite{alcaraz,schuetzdomb}. Therefore
it is not surprising that many methods for the treatment of 
equilibrium systems can be adopted\cite{alcaraz,schuetzdomb}, 
e.g.\ free fermion techniques, Bethe Ansatz, or the matrix product Ansatz.

The stationary state $|P\rangle$ reached for $t\to\infty$ is 
determined by
\begin{equation}
\label{statstate}
{H} \, |P\rangle =0,
\end{equation}
i.e.\ it corresponds to a zero-energy ground state of the Hamiltonian $H$.

In this paper we consider the following stochastic process: Particles with 
a hard core occupy the sites of a chain of length $L$. The odd sites of the
chain belong to sublattice $A$ and the even sites to sublattice $B$.
The global stochastic Hamiltonian is an alternating sum of local
two-site Hamiltonians
\begin{equation}
H=\sum_{j\in A}\mu_j h_{j,j+1}^A + \sum_{j \in B}\mu_j h_{j,j+1}^B,
\end{equation}
where $h_{j,j+1}^A$ and $h_{j,j+1}^B$ act non-trivially on sites 
$j$ and $j+1$ according to
\begin{equation}
h_{j,j+1}^A=\begin{pmatrix}   0 & 0 & 0 &0\\
                        0 & a & -\frac{1}{a} &0\\
                        0 & -a & \frac{1}{a} &0\\
                        0 & 0 & 0 &0\\
        \end{pmatrix}_{j,j+1},\qquad
h_{j,j+1}^B=\begin{pmatrix}   0 & 0 & 0 &0\\
                        0 & c & -\frac{1}{c} &0\\
                        0 & -c & \frac{1}{c} &0\\
                        0 & 0 & 0 &0\\
        \end{pmatrix}_{j,j+1}.           
\end{equation}
The local two-site basis is ordered as follows: 
$\lbrace \uparrow\uparrow,\uparrow\downarrow,\downarrow\uparrow,
\downarrow\downarrow\rbrace$ where
$\uparrow$ corresponds to an empty site and $\downarrow$ to an occupied
one. Here we have already used the spin language to emphasize the
similarities with spin systems. The only allowed processes
are jumps of particles to unoccupied neighbouring sites where the
hopping rates alternate along the bonds of the chain.

The $\mu_j\in\mathbb{R}^{+}$ are arbitrary constants which may differ
from site to site. They control the activities of the bonds and
influence dynamical properties only.


\section{CONSTRUCTION OF THE STATIONARY STATE}

Since no particles enter or leave the chain, the total particle number
$N$ is a conserved quantity and the stationary states of a chain of 
length $L$ can be classified according to the particle number $N\leq L$. 

Exact solutions of small systems  with $L\leq 4$ indicate that
the stationary state for any given particle number is unique
and, more interestingly, that this stationary state is an OGS. An OGS
is a special type of ground state of a Hamiltonian $H=\sum_{j,l} h_{jl}$, 
which is at the same time ground state of {\em all} local 
Hamiltonians $h_{jl}$. In the present case this means that all local
Hamiltonians $h_{j,j+1}^A$ and $h_{j,j+1}^B$ have to annihilate
the stationary state $|P\rangle$ since the ``ground state energy''
of $H$ is $0$ according to (\ref{statstate}).
Explicitly this has the consequence that if $| P\rangle$ has a component 
\begin{equation*}
|\alpha_1,\ldots,\alpha_{j-1},\downarrow,\uparrow,\beta_{j+2},\ldots,
\beta_L\rangle
\end{equation*}
it must also have the component  
 \begin{equation*}
|\alpha_1,\ldots,\alpha_{j-1},\uparrow,\downarrow,\beta_{j+2},
 \ldots,\beta_L\rangle
\end{equation*}
 with a relative weight given by 
\begin{equation}
\label{eq:brace1}
\left\{ \begin{matrix} a^{-2} \\ c^{-2} \end{matrix} \right\}_j :=
\begin{cases}
a^{-2} \text{ if $j$ odd},\\
c^{-2} \text{ if $j$ even}.
\end{cases}
\end{equation}

Stochastic processes on a one-dimensional lattice with a
finite interaction range and boundary interactions have a stationary
state which can be written as a matrix product state
\cite{ks,us}. This can be shown without using the explicit form of the 
boundary operators $h_1$, $h_L$. 
Therefore, reflective boundaries\footnote{In the language 
of quantum spin chains such boundaries are usually denoted open boundaries.} 
-- corresponding to vanishing boundary operators $h_1= h_L=0$
-- are allowed as well.

A suitable (grand canonical) matrix product ansatz for the stationary state 
in the presence of staggering has the form 
\begin{equation}
| P_L\rangle=\langle W| \underbrace{\mathcal{D}\otimes\bar{\mathcal{D}}
\otimes\cdots\otimes\left\{ \begin{matrix} \mathcal{D} \\ \bar{\mathcal{D}} 
\end{matrix} \right\}_L}_{\text{$L$-fold product}}|V\rangle ,
\end{equation}
where we have used (\ref{eq:brace1}) to write the state for
even and odd $L$ in a unified way and 
\begin{equation}
\mathcal{D}=    \begin{pmatrix} E\\
                                D
                \end{pmatrix},\qquad
\bar{\mathcal{D}}=      \begin{pmatrix} \bar{E}\\
                                        \bar{D}
                        \end{pmatrix}.  
\end{equation}
$E$, $D$, $\bar{E}$ and $\bar{D}$ are matrices acting on some
auxiliary vector space $A$, $|V\rangle \in A$ and $\langle W| \in
A^{\ast}$. By direct computation one verifies that this vector is a ground 
state of $H$, if one can find  matrices $X_1$, $X_2$ and $\bar{X}_1$, 
$\bar{X}_2$, which are the components of the vectors $\mathcal{X}$ and 
$\bar{\mathcal{X}}$, respectively, such that the following relations are 
fulfilled
\begin{equation}
\begin{split}
h^A(\mathcal{D}\otimes\bar{\mathcal{D}})&=\mathcal{X}\otimes\bar{\mathcal{D}}
-\mathcal{D}\otimes\bar{\mathcal{X}},\\
h^B(\bar{\mathcal{D}}\otimes\mathcal{D})&=\bar{\mathcal{X}}\otimes\mathcal{D}
-\bar{\mathcal{D}}\otimes\mathcal{X},\\
\langle W| \mathcal{X}&=0,\\
\left\{ \begin{matrix} \mathcal{X} \\ \bar{\mathcal{X}} \end{matrix} 
\right\}_L|V\rangle&=0.
\end{split}
\end{equation}

The most simple way to satisfy this algebra is given by the choice
$X_1=X_2=\bar{X}_1=\bar{X}_2=0$. Then, the only remaining equations are 
\begin{equation}
\begin{split}
h^A(\mathcal{D}\otimes\bar{\mathcal{D}})&=0,\\
h^B(\bar{\mathcal{D}}\otimes\mathcal{D})&=0
\end{split}
\end{equation}
turning $|P_L\rangle_0$ into an optimum ground state!

More explicitly, the problem of finding the ground state of $H$
becomes the problem of finding a representation of the  quadratic algebra
\begin{equation}
\begin{split}
\label{eq:alt.algebra}
E \bar{D}&=\frac{1}{a^2}D\bar{E},\\
\bar{E} D&=\frac{1}{c^2}\bar{D} E.\\
\end{split}
\end{equation}
The algebra (\ref{eq:alt.algebra}) fixes the relative weight of two
configurations in $|P_L\rangle_0$ that differ only by the interchange
of a particle and a hole at the  sites $j$ and $j+1$. This
relative weight is exactly the one that has been proposed above
by means of the optimum ground state property given by (\ref{eq:brace1}). 
Thus, in case of reflective boundaries the appearance of optimum ground 
states can be connected with the fact that the algebra necessary for the
construction of the ground state in form of a matrix product state
takes its most simple form. Nevertheless, it should be noticed that 
reflective boundaries do not enforce optimum ground states, as will be 
shown in the caveat at the end (Sec.~\ref{sec_caveat}).

Up to this point, two ways for the construction of the stationary state
of $H$ have been presented. The first way gives a recipe
for the construction of the relative probabilities of all states that
are  present in the stationary state. The
second way is connected with the matrix product approach: The ground
state problem is transformed into the purely algebraic problem of finding
the representation of the quadratic algebra (\ref{eq:alt.algebra}).  

In the following, no use of these two possibilities will be
made. Instead, a third way will be given that closely resembles the
construction in \cite{sandowschuetz}.  The starting point will be the
vacuum vector $\vac:=\prod_{i=1}^{\otimes L} |\uparrow\rangle$
describing a completely empty chain. 

Out of this vacuum all $N$-particle stationary states will be
constructed, using powers of suitable creation\footnote{Note that
  creation of particles corresponds to lowering a spin in the language
  of spin-operators.} operators. With 
\begin{equation}
q:=\begin{pmatrix} 
  1 & 0\\
  0 & a c
  \end{pmatrix}
\end{equation}
for every $j\in\lbrace 1,2,\ldots,L\rbrace$ a ``local'' particle
creation operator $b_j^-$ is given by
\begin{equation}
b_j^-=\left[\left( \frac{1}{a^2}   \right)^{\lfloor \frac{j}{2}
  \rfloor}\left( \frac{1}{c^2}   \right)^{\lfloor \frac{j-1}{2}
  \rfloor}\prod_{i=1}^{j-1}q_i\right] \cdot s_j^-.
\end{equation}
The global creation operator $B^-$ is the sum of all local creation
operators
\begin{equation}
B^-=\sum_{j=1}^{L} b_j^-.
\end{equation}
In order to shorten the notation one furthermore defines
\begin{eqnarray}
\lbrace k \rbrace_{a:c}&:=&\left( \frac{1}{a^2} \right)^{\lfloor \frac{k}{2}
  \rfloor}\left( \frac{1}{c^2}   \right)^{\lfloor \frac{k-1}{2}
  \rfloor},\quad
[k]_{a:c}:=\frac{1-(ac)^k}{1-ac},\nonumber\\ 
{[k]}_{a:c} ! &:=& [1]_{a:c}[2]_{a:c}\cdots [k]_{a:c}.
\label{eq:brace2}
\end{eqnarray}
The symbol $\lbrace k\rbrace_{a:c}$ should not be confused with
$\left\{\scriptscriptstyle  \begin{matrix} a^{-2} \\ c^{-2} \end{matrix}
\right\}_j$ defined in (\ref{eq:brace1}), which looks similar but has a 
different meaning. The floor function $\lfloor k \rfloor$ denotes the 
largest integer $\leq k$.
With these definitions the unnormalized $N$-particle stationary 
state $|N\rangle$ is built out of  the vacuum by
\begin{equation}
|N\rangle:=\frac{1}{[N]_{a:c}!}(B^-)^N\vac.
\end{equation}
In addition, one has the left complete $N$-particle state, defined as
\begin{equation}
\langle N|:=\langle \text{vac}|\frac{1}{N!}(S^+)^N.
\end{equation}
For the left complete state it is straightforward to show that
\begin{equation}
\langle N|s_k^-=\langle N-1|(1-n_k).
\end{equation}
From exact solutions of systems with $L\leq 4$ a similar relation has been 
guessed for the unnormalized $N$-particle state $|N\rangle$:
For all $k\in\lbrace 1,2,\ldots,L\rbrace$ the following identity holds
\begin{equation}
s_k^+|N\rangle=\lbrace k\rbrace_{a:c}(1-n_k)|N-1\rangle
\end{equation} 
which can be proven by direct calculation. As a consequence
one finds that the $l$-point density correlation functions obey the recursion
relation
\begin{equation}
\label{eq:central}
\langle N| n_{x_1}\ldots n_{x_l}|N\rangle=\lbrace
x_l\rbrace_{a:c}\langle N-1| n_{x_1}\ldots (1-n_{x_l})|N-1 \rangle.
\end{equation}
With this recursion relation at hand it is easy to prove that
\begin{equation}
\frac{P(\alpha_1,\ldots,\alpha_{k-1},\uparrow,\downarrow,\beta_{k+2},
\ldots,\beta_L)}{P(\alpha_1,\ldots,\alpha_{k-1},\downarrow,\uparrow,
\beta_{k+2},\ldots,\beta_L)}=\left\{ \begin{matrix} a^{-2} \\ c^{-2} 
\end{matrix}\right\}_k
\end{equation}
which suffices to show that the $N$-particle states defined above are
optimum ground states and hence
stationary states.

Note that $B^-$ does not commute with $H$ and so does {\it not} reflect 
a global symmetry. Both operators commute only, if they are restricted to 
the kernel of $H$. Therefore, the integrability found for $a=c=q$ is most 
presumably lost if $a\neq c$.


\section{DENSITY PROFILES FROM RECURSION RELATIONS}

\begin{figure}[ht]
\begin{center}
\includegraphics[width=0.85\linewidth]{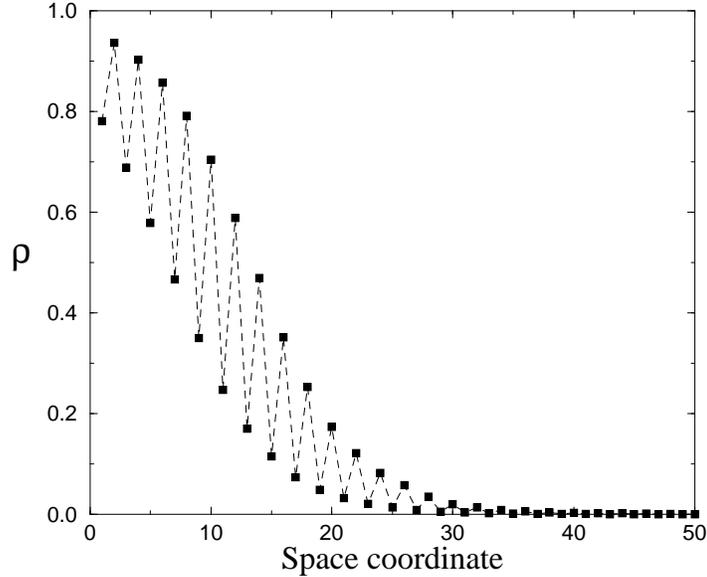}\\
\end{center}
\caption{Density profile of staggered hopping-model. $a=0.5$, $c=2.5$,
  $L=50$ and $N=10$. The dashed line corresponds to analytical data
  from (\ref{eq:prof.rec}). The boxes represent Monte Carlo data.}
\label{fig:profile1}
\end{figure}

For the calculation of density profiles, in principle, the normalization of
$|N\rangle$ has to be computed. Unfortunately, a direct computation of
this quantity is very involved and tedious. Therefore a step-by-step
method has been chosen. Assume that $Z(N-1)$ is the normalization of
$|N-1\rangle$. Assume further that $z_N$ is such that the
normalization $Z(N)$ of $|N\rangle$ is given by 
\begin{equation}
Z(N)=z_NZ(N-1).
\end{equation}
This gives with (\ref{eq:central})
\begin{equation}
\rho_k(N)=\frac{\langle N| n_k|N\rangle}{Z(N)}=
\frac{\lbrace k\rbrace_{a:c}}{z_N}(1-\rho_k(N-1)).
\end{equation}  
In the $N$-particle state naturally the identity $\sum_{k=1}^{L} \rho_k(N)=N$
holds. Thus,
\begin{equation}
z_N=\frac{1}{N}\sum_{k=1}^{L}\lbrace k\rbrace_{a:c}(1-\rho_k(N-1)).
\end{equation}

\begin{figure}[ht]
\begin{center}
\includegraphics[width=0.85\linewidth]{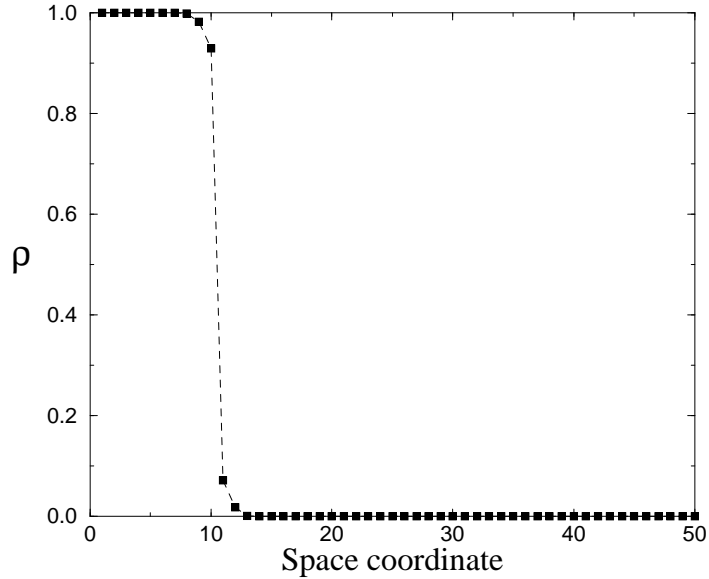}\\
\end{center}
\caption{Density profile of staggered hopping-model. $a=2.0$, $c=4.0$,
  $L=50$ and $N=10$. The dashed line corresponds to analytical data
  from (\ref{eq:prof.rec}). The boxes represent Monte Carlo data.}
\label{fig:profile2}
\end{figure}

Finally, one finds a recursion relation for the local density of an
$N$-particle stationary state
\begin{equation}
\label{eq:prof.rec}
\rho_k(N)=N\frac{\lbrace k\rbrace_{a:c}(1-\rho_k(N-1))}{\sum_{j=1}^{L}
\lbrace j\rbrace_{a:c}(1-\rho_j(N-1))}.
\end{equation}
Starting from the normalized stationary state $\vac$,  which belongs to
$N=0$ and has $\rho_k(0)=0$, the density profiles for all $N$ can be
computed. The results have been compared with Monte Carlo data (see
Fig.~\ref{fig:profile1} and Fig.~\ref{fig:profile2}) showing
perfect agreement. 
Generically we find exponentially decaying density profiles.
If one of the parameters is larger than one, whereas the
other parameter is smaller than one,  the  density profiles
clearly reflect the sublattice structure of the Hamiltonian, since the 
slope of the profile is
alternatingly positive and negative along the bonds (see
Fig.~\ref{fig:profile1}). 
The shape of the density profile can be explained 
qualitatively in the following way: on the odd bonds hopping to the 
right is favoured because $a^{-1}=2$ and $a=1/2$.
As a consequence, the density at the left end of an odd bond
is smaller than the density at the right end. Along the even bonds one
has the opposite scenario since $c=5/2$ and $c^{-1}=2/5$. Finally, the
high density in the left part of the system is due to the fact that
the maximal hopping amplitude can be found at the even bonds which
favour hopping to the left.

If both parameters $a$ and $c$ are larger or smaller than one, we
find a different behaviour. Now the density profiles become monotonous 
functions of $\rho$ and the sublattice structure is hidden (see
Fig.~\ref{fig:profile2}).


\section{A CAVEAT}
\label{sec_caveat}

As presented above, optimum ground states come into play naturally if
one encounters stochastic processes with reflective
boundaries. Therefore one might conjecture that the stationary state
of such processes is always of this special type. However, this is not
true. In the following a $\mathbb{Z}_3$-symmetric reaction-diffusion
model will be presented, which has zero-energy ground states that are
not (always) optimum ground states.

The sites of a chain of length $L$  are coloured with the elements of
$\mathbb{Z}_3$, i.e., one has three types of particles ($k=0,1,2$) on 
a completely filled chain. Two particles on neighbouring sites react 
according to the following rules (algebraic manipulations are done 
according to $\mathbb{Z}_3$) 
\begin{equation}
\begin{split}
(k,k+1) &\rightarrow (k+1,k)\phantom{k+1}\qquad\text{ with rate }\alpha,\\
(k,k+2) &\rightarrow (k+1,k+1)\phantom{k}\qquad\text{ with rate }\beta. 
\end{split}
\end{equation}
Thus, the local Hamiltonian is given by 
\begin{equation}
h={\scriptscriptstyle 
\begin{pmatrix}
  0 & 0 & 0 & 0 & 0 & 0 & 0 & -\beta & 0 \\
  0 & \alpha & 0 & 0 & 0 & 0 & 0 & 0  & 0 \\
  0 & 0 & \beta & 0 & 0 & 0 & -\alpha & 0 & 0 \\
  0 & -\alpha & 0 & \beta & 0 & 0 & 0 & 0 & 0 \\
  0 & 0 & -\beta & 0 & 0 & 0 & 0 & 0 & 0 \\
  0 & 0 & 0 & 0 & 0 & \alpha & 0 & 0 & 0 \\
  0 & 0 & 0 & 0 & 0 & 0 & \alpha & 0 & 0 \\
  0 & 0 & 0 & 0 & 0 & -\alpha & 0 & \beta & 0 \\
  0 & 0 & 0 &  -\beta & 0 & 0 & 0 & 0 & 0  
\end{pmatrix}}.
\end{equation}
The operators $h_{j,j+1}$ act as $h$ on sites $j,j+1$ and as identity 
on the remaining sites of the chain, and hence  the global stochastic 
Hamiltonian reads
\begin{equation}
H=\sum_{j=1}^{L-1}h_{j,j+1}.
\end{equation}
The exact solution for $L=2$ shows a 
threefold degenerate ground state
\begin{equation}
|\psi\rangle_2=|k,k\rangle \text{ with } k\in\mathbb{Z}_3.
\end{equation}
These states are optimum ground states. 

Looking at $L=3$ a new situation arises. One has again a threefold
optimum ground state
\begin{equation}
|\psi\rangle_3=|k,k,k\rangle \text{ with } k\in\mathbb{Z}_3.
\end{equation}
In addition, one finds two qualitatively different ground states 
\begin{equation}
\begin{split}
|\varphi_1\rangle_3&=\sum_{k\in\mathbb{Z}_3}\left[|k,k,k+2\rangle+
\frac{\beta}{\alpha}|k,k+1,k+1\rangle+
\frac{\beta}{\alpha+\beta}|k,k+2,k\rangle\right],\\
|\varphi_2\rangle_3&=\sum_{k\in\mathbb{Z}_3}\left[|k,k+2,k+2\rangle+
\frac{\beta}{\alpha}|k,k,k+1\rangle+
\frac{\beta}{\alpha+\beta}|k,k+1,k\rangle\right].\\
\end{split}
\end{equation}
These states are not optimum ground states. Moreover 
\begin{equation}
\label{eq:nonopt}
h_{1,2}|\varphi_j\rangle_3=-h_{2,3}|\varphi_j\rangle_3 \neq 0
\qquad \text{ with } j=1,2.
\end{equation}
The occurrence of such states  
might be connected with the existence of an additional symmetry.

The last result can also be understood in the context of MPA. As the
two additional states are not optimum ground states, the cancelling vector
$\mathcal{X}$ has to be nontrivial. Therefore, 
\begin{equation}
|\varphi_j\rangle_3=\langle W|\mathcal{D}\otimes\mathcal{D}
\otimes\mathcal{D}|V\rangle. 
\end{equation}  
Using the fact that $\langle W|$ and $|V\rangle$ must be in the kernel
of $\mathcal{X}$ one finds
\begin{equation}
\begin{split}
h_{1,2}\langle
W|\mathcal{D}\otimes\mathcal{D}\otimes\mathcal{D}|V\rangle&=-\langle
W|\mathcal{D}\otimes\mathcal{X}\otimes\mathcal{D}|V\rangle,\\
h_{2,3}W|\mathcal{D}\otimes\mathcal{D}\otimes\mathcal{D}|V\rangle&=\langle
W|\mathcal{D}\otimes\mathcal{X}\otimes\mathcal{D}|V\rangle,
\end{split}
\end{equation}
and thus (\ref{eq:nonopt}) holds.

Most interestingly systems of length $L=4$ and $L=5$ -- like the
two-site system -- possess only
threefold degenerate optimum ground states. But then, for $L=6$,  two
additional  ground states
show up again. 
This might be related to an additional symmetry occuring for
system sizes which are an integer multiple of three.

This example shows that reflective boundary conditions do not imply
the existence of an OGS. However, it still leaves open the possibility
that systems with reflecting boundary conditions and a {\em unique} stationary 
state have optimum ground states.


\section{CONCLUSIONS}

We have presented the exact solution for the stationary state of
a stochastic process with staggered hopping rates.
Using the similarity of the quantum formalism to the  master equation
the model could be interpreted as a spin chain. Its ground state
-- corresponding to the stationary state of the stochastic process --
was shown to be an optimum ground state, i.e.\ it is at the
same time ground state of all local Hamiltonians $h_{j,j+1}$
which make up $H$. Furthermore the $l$-point density correlation functions
have been determined recursively.

Apart from the fact that the model is one of the very few exactly solvable
models with staggered interactions it might have practical applications.
A similar model has been proposed by Kolomeisky and Widom\cite{kolo}.
It is a simplified ratchet model (for a recent review, see \cite{reimann})
for molecular motors effectively describing the motion of a motor
protein on a micro-tubule\cite{juelicher}.


\section*{ACKNOWLEDGEMENTS}
This work was performed within the research program of the
Sonderforschungsbereich 341, K\"oln-Aachen-J\"ulich.


\end{document}